\documentclass{aastex}
\usepackage{spr-astr-addons}
\usepackage{url}
\RequirePackage{color}

\begin{document}

\title{A cosmological scenario from the Starobinsky model within the $f(R,T)$ formalism}
\slugcomment{Not to appear in Nonlearned J., 45.}
\shorttitle{Short article title}
\shortauthors{Autors et al.}

\author{P.H.R.S. Moraes} 
\affil{UNINA - Universit\'a degli Studi di Napoli Federico II, Dipartamento di Fisica - Napoli I-80126, Italy}
\affil{ITA - Instituto Tecnol\'ogico de Aeron\'autica - Departamento de F\'isica, 12228-900, S\~ao Jos\'e dos Campos, S\~ao Paulo, Brazil}
\email{moraes.phrs@gmail.com}
\and
\author{P.K. Sahoo}
\affil{Department of Mathematics, Birla Institute of Technology and Science - Pilani, Hyderabad Campus, Hyderabad-500078, India}
\email{pksahoo@hyderabad.bits-pilani.ac.in}
\and
\author{G. Ribeiro}
\affil{UNESP - Universidade Estadual Paulista ``J\'ulio de Mesquita Filho'' - Departamento de F\'isica e Qu\'imica, 12516-410, Guaratinguet\'a, S\~ao Paulo, Brazil}
\email{ribeiro.gabriel.fis@hotmail.com}
\and
\author{R.A.C. Correa}
\affil{SISSA - Scuola Internazionale Superiore di Studi Avanzati, via Bonomea, 265, I-34136 Trieste, Italy}
\affil{ITA - Instituto Tecnol\'ogico de Aeron\'autica - Departamento de F\'isica, 12228-900, S\~ao Jos\'e dos Campos, S\~ao Paulo, Brazil}
\email{fis04132@gmail.com}

\begin{abstract}
In this paper we derive a novel cosmological model from the $f(R,T)$ theory of
gravitation, for which $R$ is the Ricci scalar and $T$ is the trace of the
energy-momentum tensor. We consider the functional form $f(R,T)=f(R)+f(T)$, with $f(R)$ being
the Starobinksy model, named $R+\alpha R^{2}$, and $f(T)=2\gamma T$, with $\alpha$
and $\gamma$ being constants. We show that a hybrid expansion law form for the scale factor is a solution for the derived Friedmann-like equations. In this way, the model is able to predict both the decelerated and the accelerated regimes of expansion of the universe, with the transition redshift between these stages being in accordance with recent observations. We also apply the energy conditions to our material content solutions. Such an application makes us able to obtain the range of acceptability for the free parameters of the model, named $\alpha$ and $\gamma$. 
\end{abstract}


\section{Introduction}

\label{sec:int}

The $f(R)$ theories of gravity (\cite{sotiriou/2010,de_felice/2010}) are an
optimistic alternative to the shortcomings General Relativity (GR) faces as
the underlying gravitational theory, such as those discussed by \cite%
{padmanabhan/2003,antoniadis/2013,demorest/2010,bull/2016}. They can account
for the cosmic acceleration (\cite{riess/1998,perlmutter/1999}) with no need
for a cosmological constant, providing good match between theory and
cosmological observations (\cite{tsujikawa/2008,capozziello/2005,nojiri/2006}).

Particularly, in standard cosmology, derived from GR, the dark energy and
dark matter should compose $\sim95\%$ of the universe (\cite{hinshaw/2013}),
but their nature is still dubious (\cite%
{dil/2017,behrouz/2017,germani/2017,jennen/2016,evslin/2016,liu/2016,rinaldi/2017,zhang/2017,albert/2017}).

Another crucial trouble surrounding GR is the difficulty in quantizing it.
Attempts to do so have been proposed (\cite%
{fradkin/1985,witten/1986,friedan/1986}) and can, in future, provide us a
robust and trustworthy model of gravity - quantum mechanics unification.

Meanwhile it is worthwhile to attempt to consider the presence of quantum
effects in gravitational theories. Those effects can rise from the
consideration of terms proportional to the trace of the energy-momentum
tensor $T$ in the gravitational part of the $f(R)$ action, yielding the $%
f(R,T)$ gravity theories (\cite{harko/2011}). Those theories were also
motivated by the fact that although $f(R)$ gravity is well behaved in
cosmological scales, the Solar System regime seems to rule out most of the $%
f(R)$ models proposed so far (\cite%
{erickcek/2006,chiba/2007,capozziello/2007b,olmo/2007}). Furthermore,
rotation curves of spiral galaxies were constructed in $f(R)$ gravity, but
the results did not favour the theory, as it can be checked in \cite%
{chiba/2003,dolgov/2003,olmo/2005}. The structure and cosmological properties of the modified gravity starting from $f(R)$ theory to power-counting renormalizable covariant gravity were presented by \cite%
{nojiri/2011}. The review in \cite%
{nojiri/2017} describes the cosmological developments regarding inflation, bounce and late time evolution in $f(R)$,  $f(\mathcal{G})$ and $f(\mathcal{T})$ modified theories of gravity, with $\mathcal{G}$ and $\mathcal{T}$ being the Gauss-Bonnet and torsion scalars.

Despite its recent elaboration, $f(R,T)$ gravity has already been applied to
a number of areas, such as Cosmology (\cite%
{moraes/2015,mrc/2016,ms/2016,ms/2017,myrzakulov/2012,reddysn/2012,singh/2015}
and Astrophysics (\cite%
{mam/2016,amam/2016,ms/2017b,mcl/2017,das/2016,sy/2014,yousaf/2017b,baffou/2017}).
Particularly, solar system tests have been applied to $f(R,T)$ gravity (\cite%
{shabani/2014}) and the dark matter issue was analysed by \cite%
{zaregonbadi/2016}. The late time behaviour of cosmic fluids consisting of collisional self-interacting dark matter and radiation was discussed by \cite%
{zubair/2018}. Moreover, considering the metric and the affine connection as independent field variables, the Palatini formulation of the $f(R,T)$ gravity can be seen in (\cite%
{wu/2018,barrientos/2018}).

By investigating the features of an $f(R,T)$ or $f(R)$ model, one realizes
the strong relation they have with the functional form of the chosen
functions for $f(R,T)$ and $f(R)$, as well as with their free parameters
values. In fact, a reliable method to constraint those ``free" parameters to
values that yield realistic models can be seen in \cite{cm/2016} and \cite%
{cmsdr/2015} for these theories, respectively.

In the $f(R)$ gravity, a reliable and reputed functional form was proposed by
A.A. Starobinsky as (\cite{starobinsky/1980,starobinsky/2007})

\begin{equation}  \label{i1}
f(R)=R+\alpha R^{2},
\end{equation}
which is known as Starobinsky Model (SM), with $\alpha$ a constant. It
predicts a quadratic correction of the Ricci scalar to be inserted in the
gravitational part of the Einstein-Hilbert action.

SM has been deeply applied to the cosmological and astrophysical contexts in the
literature. Starobinsky showed that a cosmological model obtained from Eq.(%
\ref{i1}) can satisfy cosmological observational tests (\cite%
{starobinsky/2007}). On the other hand, the model seems to predict an
overproduction of scalarons in the very early universe. In an astrophysical context, SM is also of great importance. In \cite%
{sharif/2017}, the authors have explored the source of a gravitational
radiation in SM by considering axially symmetric dissipative dust under
geodesic condition. In \cite{resco/2016}, it has been shown that in SM it is
possible to find neutron stars with $2M_{\odot}$, which raises as an
important alternative to some of the GR shortcomings mentioned above (\cite%
{antoniadis/2013,demorest/2010}). In \cite{astashenok/2015}, the
macroscopical features of quark stars were obtained.

Our proposal in this paper is to construct a cosmological scenario from an $%
f(R,T)$ functional form whose $R-$dependence is the same as in the SM, i.e.,
with a quadratic extra contribution of $R$, as in Eq.(\ref{i1}). The $T-$%
dependence will be considered to be linear, as $2\gamma T$, with $\gamma$ a
constant. Therefore, we will take

\begin{equation}  \label{i2}
f(R,T)=R+\alpha R^{2}+ 2\gamma T.
\end{equation}

As far as the present authors know, the SM has not been considered for the $%
R-$dependence within $f(R,T)$ models for cosmological purposes so far, only
in the study of astrophysical compact objects (\cite%
{zubair/2015,noureen/2015,noureen/2015b}) and wormholes (\cite{zubair/2016,yousaf/2017}). We believe this is due to the
expected high non-linearity of the resulting differential equation for the
scale factor. Anyhow, the consideration of linear material corrections
together with quadratic geometrical terms can imply interesting outcomes in
a cosmological perspective as it does in the astrophysical level. 

\section{The $f(R,T)=R+\protect\alpha R^{2}+ 2\protect\gamma T$ gravity}

\label{sec:frtg}

Following the steps in \cite{harko/2011}, we can write the $f(R,T)=R+\alpha
R^{2}+ 2\gamma T$ gravity total action as

\begin{equation}  \label{frt1}
S=\frac{1}{16\pi}\int (R+\alpha R^{2}+ 2\gamma T)\sqrt{-g}d^4x+\int L_m\sqrt{%
-g}d^{4}x,
\end{equation}
in which $g$ is the determinant of the metric $g_{\mu\nu}$, $L_m$ is the
matter lagrangian and we are working with natural units.

By taking $L_m=-p$, with $p$ being the pressure of the universe, the
variational principle applied in Eq.(\ref{frt1}) yields the following field
equations:

\begin{multline}  \label{frt2}
\frac{1}{2}G_{\mu\nu}+\alpha\left(R_{\mu\nu}-\frac{1}{4}Rg_{\mu\nu}+g_{\mu%
\nu}\Box-\nabla_\mu\nabla_\nu\right)R\\=4\pi T_{\mu\nu}+\gamma\left[%
T_{\mu\nu}+\left(p+\frac{1}{2}T\right)g_{\mu\nu}\right].
\end{multline}

In (\ref{frt2}), $G_{\mu\nu}$ is the Einstein tensor, $R_{\mu\nu}$ is the
Ricci tensor, $T_{\mu\nu}=\text{diag}(\rho,-p,-p,-p)$, $\rho$ is the matter-energy
density of the universe and $T=\rho-3p$. Still in (\ref{frt2}), it can be
straightforwardly seen that the limit $\alpha=\gamma=0$ recovers the GR
field equations.

Also, the above choice for the matter lagrangian is usually assumed in the
literature as it can be checked in \cite{harko/2011,mam/2016,mcl/2017},
among many others.

\section{The $f(R,T)=R+\protect\alpha R^{2}+ 2\protect\gamma T$ cosmology}

\label{sec:frtc}

Let us assume a flat Friedmann-Robertson-Walker metric in the field equations above. Such a substitution yields the following
Friedmann-like equations

\begin{equation}
\left( \frac{\dot{a}}{a}\right) ^{2}+6\alpha G(a,\dot{a},\ddot{a},\dot{\ddot{%
a}})=\frac{8\pi }{3}\rho +\gamma \left( \rho -\frac{p}{3}\right) ,  \label{00}
\end{equation}%
\begin{equation}
\frac{\ddot{a}}{a}+\frac{1}{2}\left( \frac{\dot{a}}{a}\right) ^{2}+6\alpha
S(a,\dot{a},\ddot{a},\dot{\ddot{a}},\ddot{\ddot{a}})=-4\pi p+\frac{1}{2}%
\gamma (\rho -p),  \label{11}
\end{equation}

\noindent where we are using the following definitions%
\begin{multline}
G(a,\dot{a},\ddot{a},\dot{\ddot{a}}) \equiv 2\left( \frac{\dot{a}}{a}%
\right) ^{2}\left[ \frac{\ddot{a}}{a}+\frac{\dot{\ddot{a}}}{a}-\frac{3}{2}%
\left( \frac{\dot{a}}{a}\right) ^{2}\right] -\left( \frac{\ddot{a}}{a}%
\right) ^{2},  \label{ad1} 
\end{multline}

\begin{multline}
S(a,\dot{a},\ddot{a},\dot{\ddot{a}},\ddot{\ddot{a}})\\ \equiv \frac{3}{2}%
\left[ \left( \frac{\dot{a}}{a}\right) ^{4}+\frac{\ddot{a}}{a}\right] +2%
\left[ \frac{\dot{a}\dot{\ddot{a}}}{a^{2}}-3\left( \frac{\dot{a}}{a}\right)
^{2}\frac{\ddot{a}}{a}\right] +\frac{\ddot{\ddot{a}}}{a}.  \label{ad2}
\end{multline}

In the equations above, $a=a(t)$ is the scale factor and dots represent time
derivatives. Once again, the limit $\alpha =\gamma =0$ retrieves the
standard formalism.

In terms of the Hubble parameter $H=\frac{\dot{a}}{a}$, the values of $\rho$
and $p$ from Equations (\ref{00})-(\ref{11}) are

\begin{equation}
\rho =\frac{F(\gamma )}{2}[\gamma \tilde{F}(H)-(8\pi +\gamma )\tilde{G}(H)],
\label{12}
\end{equation}

\begin{equation}
p=F(\gamma )[4\pi \bar{F}(H)+3\gamma \bar{G}(H)],  \label{13}
\end{equation}

\noindent with the following definitions%
\begin{equation}
\tilde{F}(H) \equiv -F_{4}(H)-6\alpha[-3F_{2}(H)+G_{3}(H)+3F_{1}(H)] , 
\end{equation}
\begin{equation}
\tilde{G}(H) \equiv 3\left\{ H^{2}+6[G_{4}(H)+F_{3}(H)]\alpha \right\} , 
\end{equation}
\begin{equation}
\bar{F}(H) \equiv F_{4}(H)+18\alpha F_{2}(H)-F_{1}(H)+\dot{H}+2\dot{H}^{2},
\end{equation}
\begin{equation}
\bar{G}(H) \equiv -H^{2}+\dot{H}+3\alpha [G_{1}(H)+G_{2}(H)+2H^{5}+6H^{3}%
\dot{H}].
\end{equation}
For Equations (\ref{12})-(\ref{13}), we considered \\ $F(\gamma )=\frac{1}{(32\pi ^{2}+16\pi \gamma +\gamma ^{2})}$. Moreover, $F_{i}$ and $G_{i}$, with $i=1,2,3,4$, are
functions of $H$ and its time derivatives, expressed by the following:

\begin{equation}
F_{1}(H) = H^{2}\left( 1+4\dot{H}\right) , 
\end{equation}
\begin{equation}
F_{2}(H) = H^{4}-4H\ddot{H}, 
\end{equation}
\begin{equation}
F_{3}(H) = 2\left( H^{5}+3H^{3}\dot{H}\right) , 
\end{equation}
\begin{equation}
F_{4}(H) = 3H^{2}+2\dot{H},
\end{equation}
\begin{equation}
G_{1}(H) = H^{4}+\dot{H}\left( 3+7\dot{H}\right) , 
\end{equation}
\begin{equation}
G_{2}(H) = -12H\ddot{H}+H^{2}\left( -3-12\dot{H}+2\ddot{H}\right) -2\dot{%
\ddot{H}},
\end{equation}
\begin{equation}
G_{3}(H) = 3\dot{H}\left( 1+2\dot{H}\right) +2\dot{\ddot{H}}, 
\end{equation}
\begin{equation}
G_{4}(H) = -2H^{4}-\dot{H}^{2}+2H^{2}\ddot{H}.
\end{equation}

We can consider, as a solution for Eqs.(\ref{12})-(\ref{13}), the scale
factor in the hybrid expansion law form (\cite{akarsu/2014}):

\begin{equation}  \label{14}
a(t)=e^{mt}t^n,
\end{equation}
with $m$ and $n$ being constants. It can be seen that such a form for the
scale factor consists of a product of power law and exponential law
functions. Eq.(\ref{14}) mimics the power law and de-Sitter cosmologies as
particular cases and can predict the transition from a decelerated to an
accelerated regime of the universe expansion. It has been applied to
Brans-Dicke models in \cite{akarsu/2014}, yielding observational constraints
to $m$ and $n$.

From (\ref{14}), the Hubble and deceleration parameters are

\begin{equation}  \label{15}
H=m+\frac{n}{t},
\end{equation}
\begin{equation}  \label{16}
q=-\frac{\ddot{a}}{aH^2}=-1+\frac{n}{(mt+n)^2},
\end{equation}
such that the deceleration parameter is defined in such a way that its
negative values describe an accelerated expansion of the universe. 

We know that the universe not always has accelerated its expansion (\cite{riess/1998,perlmutter/1999}). The accelerated regime of expansion is considered to be a late-time phenomenon. 

In this way, one can choose the constants $m$ and $n$ in such a way that the power-law dominates over
exponential law in the early universe and the exponential law dominates over
power-law at late times. Therefore, the decelerated and
accelerated regimes of the universe expansion can be respectively well described, as well as the transition between these regimes.

From (\ref{16}) it is clear that there is a transition from deceleration to
acceleration phases of the universe expansion at $t=\frac{1}{m}(-n\pm\sqrt{n}%
)$, with $0<n<1$. Since the negativity of the second term leads to a
negative time, which indicates an non-physical situation, we conclude that
the cosmic transition may have occurred at $t=\frac{\sqrt{n}-n}{m}$.

From $\frac{a(t)}{a_0}=\frac{1}{1+z}$, with $a_0=1$ being the present value
of the scale factor and $z$ being the redshift, we obtain the following
time-redshift relation:

\begin{equation}  \label{17}
t=\frac{n}{m}W\left[\frac{m}{n} \left(\frac{1}{1+z}\right)^{1/n}\right],
\end{equation}
where $W$ denotes the Lambert function (also known as ``product logarithm'').

Using Equation (\ref{17}), we can plot the deceleration parameter with
respect to the redshift, which can be appreciated in Fig.\ref{fig2} below, in which the values chosen for $m$ and $n$ are in
agreement with the observational constraints found in \cite{akarsu/2014}.

\begin{figure}[t]
\centering
\includegraphics[width=80mm]{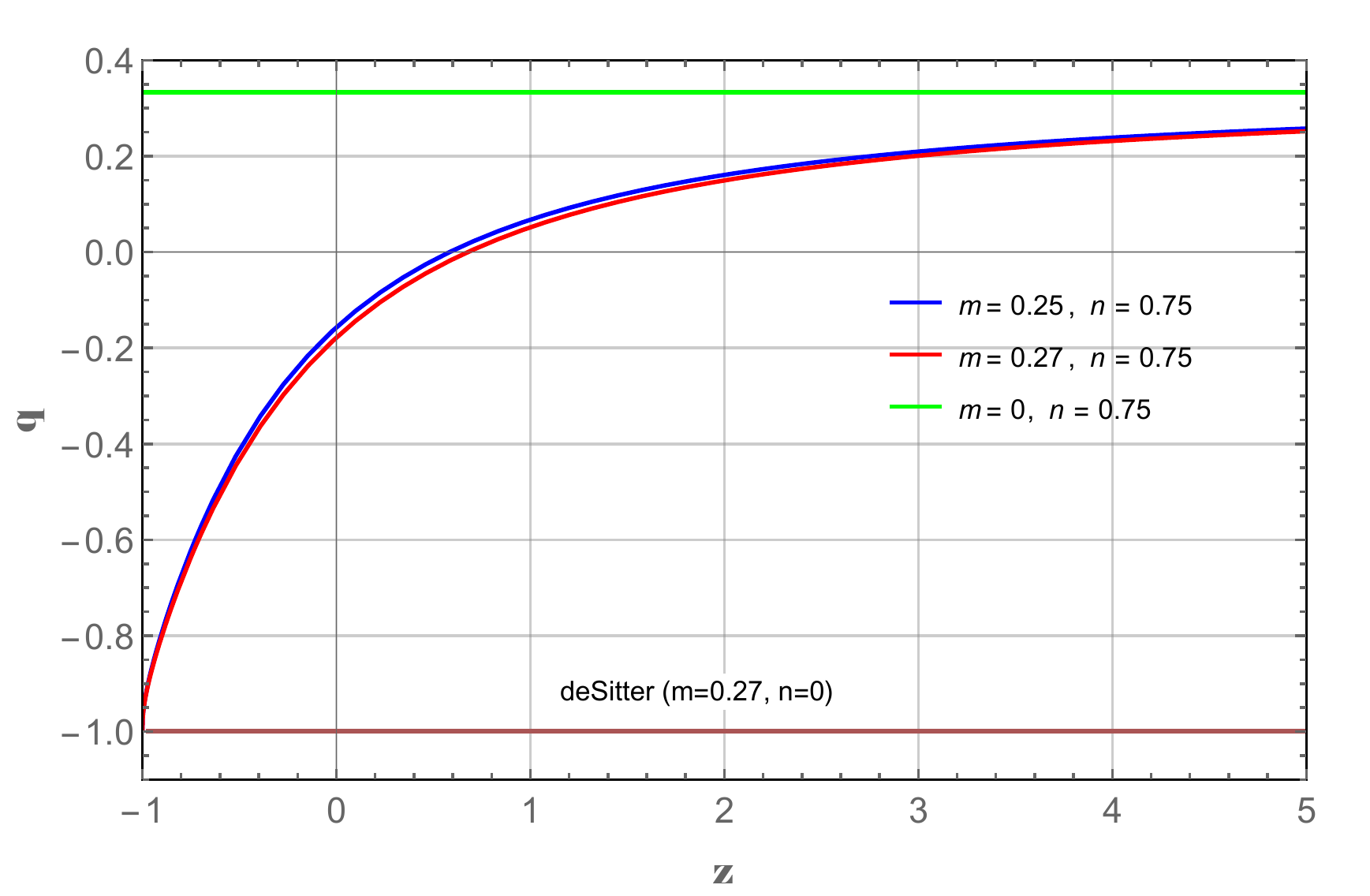}
\caption{Deceleration parameter evolution in redshift for different values
of $m$ and $n$.}\label{fig2}
\end{figure}

Plotting $q$ as a redshift function has the advantage of checking the
reliability of the model, through the redshift value in which the
decelerated-accelerated expansion of the universe transition occurs. We will
call the transition redshift as $z_{tr}$ and in our model it can be seen
that it depends directly on the parameter $m$. From Fig.1, the
transition occurs at $z_{tr} =0.5836,0.6777$, corresponding to $m=0.25,0.27$%
, respectively. The values of the transition redshift $z_{tr}$ for our model
model are in accordance with the observational data, as one can check in 
\cite{Capozziello14,Capozziello15,Farooq17}.

Now, let us write the solutions for the material content of our model, named $\rho$ and $p$. Using (\ref{15}) in Eqs.(\ref{12})-(\ref{13}), we have

\begin{equation}
\rho =\frac{F(\gamma )}{2}[-\gamma (\mathcal{P}_{1}+\mathcal{P}_{2})+%
\mathcal{P}_{3}+3(8\pi +\gamma )(\mathcal{P}_{4}+\mathcal{P}_{5}+\mathcal{P}%
_{6})],  \label{rhot}
\end{equation}

\begin{equation}
p=F(\gamma )[-(\mathcal{G}_{1}+\mathcal{G}_{2}+\mathcal{G}_{3}+%
\mathcal{G}_{4})+\mathcal{G}_{5}],
\end{equation}

\noindent where we are using the definitions%
\begin{multline}
\mathcal{P}_{1} \equiv 32\alpha n\mathcal{F}_{3}(t)^{3} \times \\ \left\{ 4\mathcal{F}_{1}(t)\left[ 1-%
\frac{\mathcal{F}_{1}(t)}{2}\right] +n\left[ 1-\frac{\mathcal{F}_{3}(t)^{-2}}{2}\right]
-2\right\} , 
\end{multline}
\begin{equation}
\mathcal{P}_{2} \equiv 18\alpha \mathcal{F}_{2}(t)^{4}\left[ \mathcal{F}_{2}(t)^{-2}-1\right] ,
\end{equation}
\begin{equation}
\mathcal{P}_{3} \equiv 3\mathcal{F}_{2}(t)^{2}-2n\mathcal{F}_{3}(t)^{2}, \\
\end{equation}
\begin{equation}
\mathcal{P}_{4} \equiv 6\alpha n\mathcal{F}_{3}(t)^{5}\left\{ 2\mathcal{F}_{1}(t)^{2}\left[
2-3\mathcal{F}_{1}(t)\right] -n\mathcal{F}_{3}(t)^{-1}\right\} , 
\end{equation}
\begin{equation}
\mathcal{P}_{5} \equiv 12\alpha \mathcal{F}_{2}(t)^{5}\left[ 1-\mathcal{F}_{2}(t)^{-1}\right] ,
\end{equation}
\begin{multline}
\mathcal{G}_{1} \equiv \alpha n\mathcal{F}_{3}(t)^{4}\mathcal{F}_{1}(t)^{2}\{9\gamma
\mathcal{F}_{3}(t)\left[ 6\mathcal{F}_{1}(t)-4\right] \\ -2\mathcal{F}_{1}^{-1}(t)G(\gamma )+G(\gamma )\},
\end{multline}
\begin{equation}
\mathcal{G}_{2} \equiv n\mathcal{F}_{3}(t)^{2}\left[ \frac{\alpha }{4}G(\gamma
)-3\gamma -8\pi \right] , 
\end{equation}
\begin{equation}
\mathcal{G}_{3} \equiv \alpha n^{2}\mathcal{F}_{3}(t)^{2}\left[ G(\gamma )+18\gamma
\right] +\alpha G(\gamma ), 
\end{equation}
\begin{equation}
\mathcal{G}_{4} \equiv \alpha n\mathcal{F}_{3}(t)^{4}G(\gamma ), 
\end{equation}
\begin{multline}
\mathcal{G}_{5} \equiv \mathcal{F}_{2}(t)^{2} \times \\ \left\{ 9\alpha \mathcal{F}_{2}(t)^{2}\left[
-2\gamma \mathcal{F}_{2}(t)-\gamma -8\pi \right] -\frac{\alpha }{4}G(\gamma )+3\gamma
+12\pi \right\},
\end{multline}
with $G(\gamma)=-108\gamma -288\pi$ and $\mathcal{F}_{j}$, with $%
j=1,2,3$, being functions of $t$ only, defined as:

\begin{eqnarray}
\mathcal{F}_{1}(t)=mt+n, \\
\mathcal{F}_{2}(t)=m+\frac{n}{t},  \\
\mathcal{F}_{3}(t)=\frac{1}{t} \label{um}.  
\end{eqnarray}

\begin{figure}[h]
\centering
\includegraphics[width=80mm]{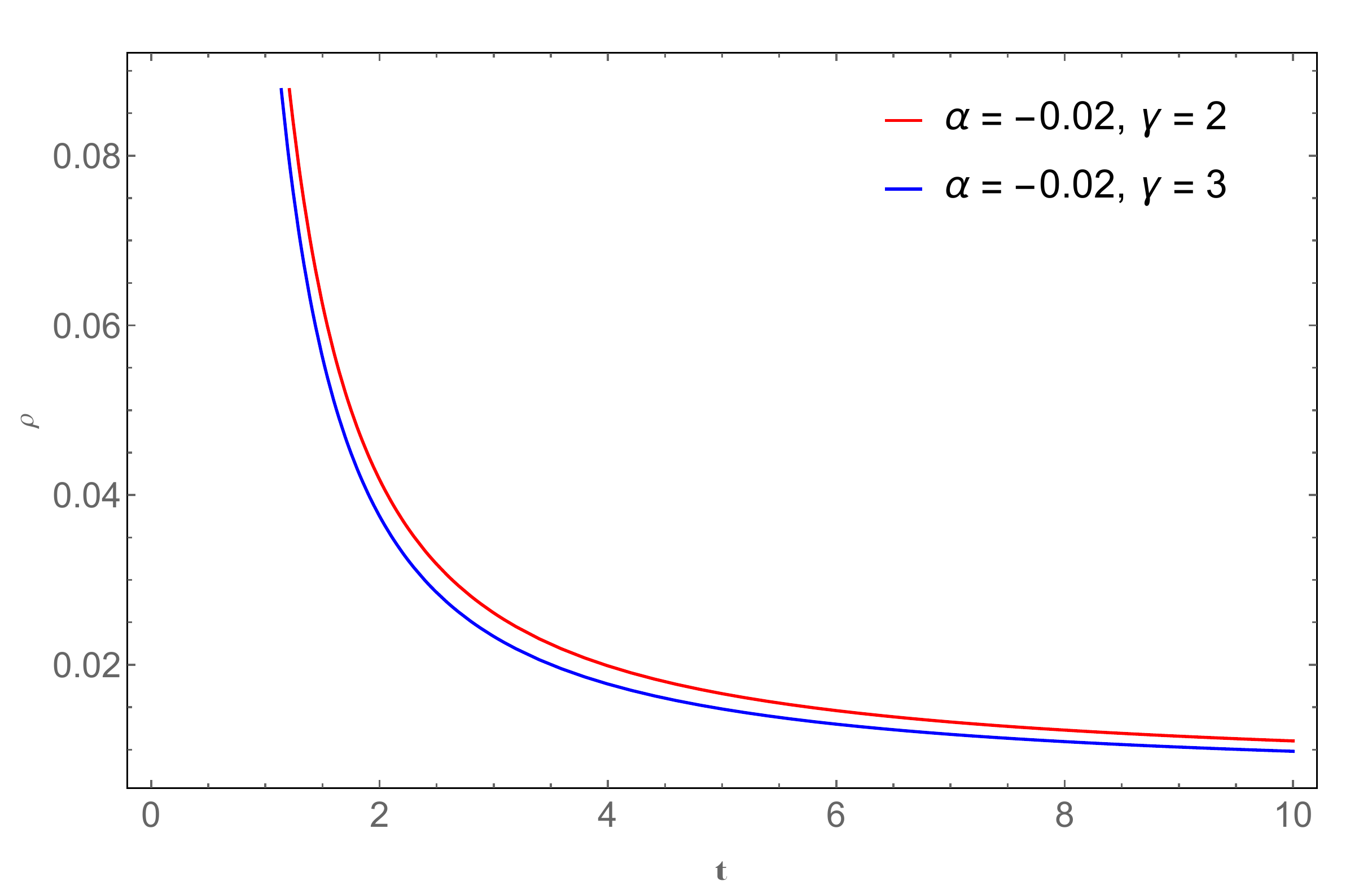}
\caption{Evolution of $\protect\rho$ with time.}\label{fig3}
\end{figure}

\begin{figure}[h]
\centering
\includegraphics[width=80mm]{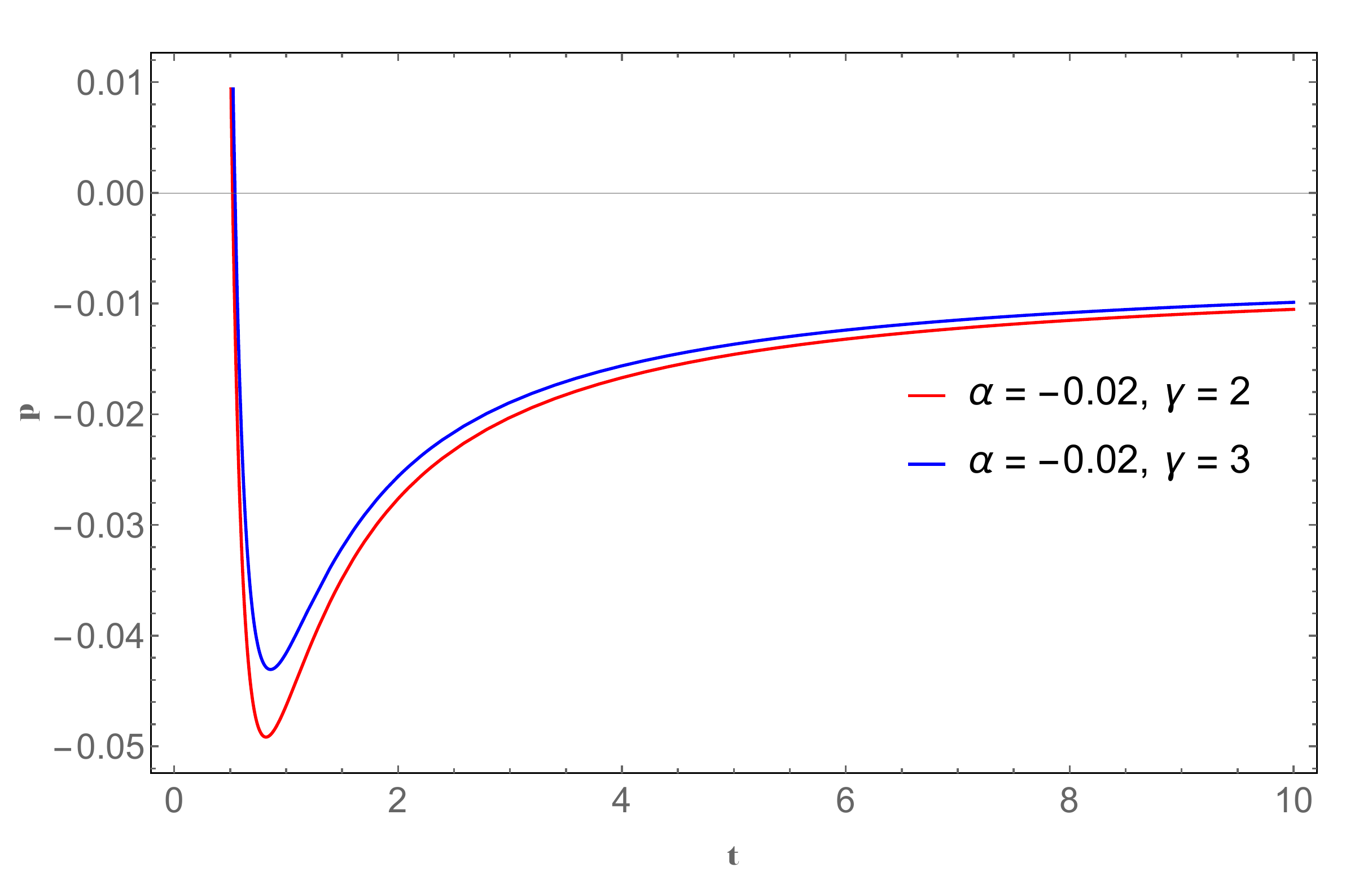}
\caption{Evolution of $p$ with time.}\label{fig4}
\end{figure}

\begin{figure}[h]
\centering
\includegraphics[width=80mm]{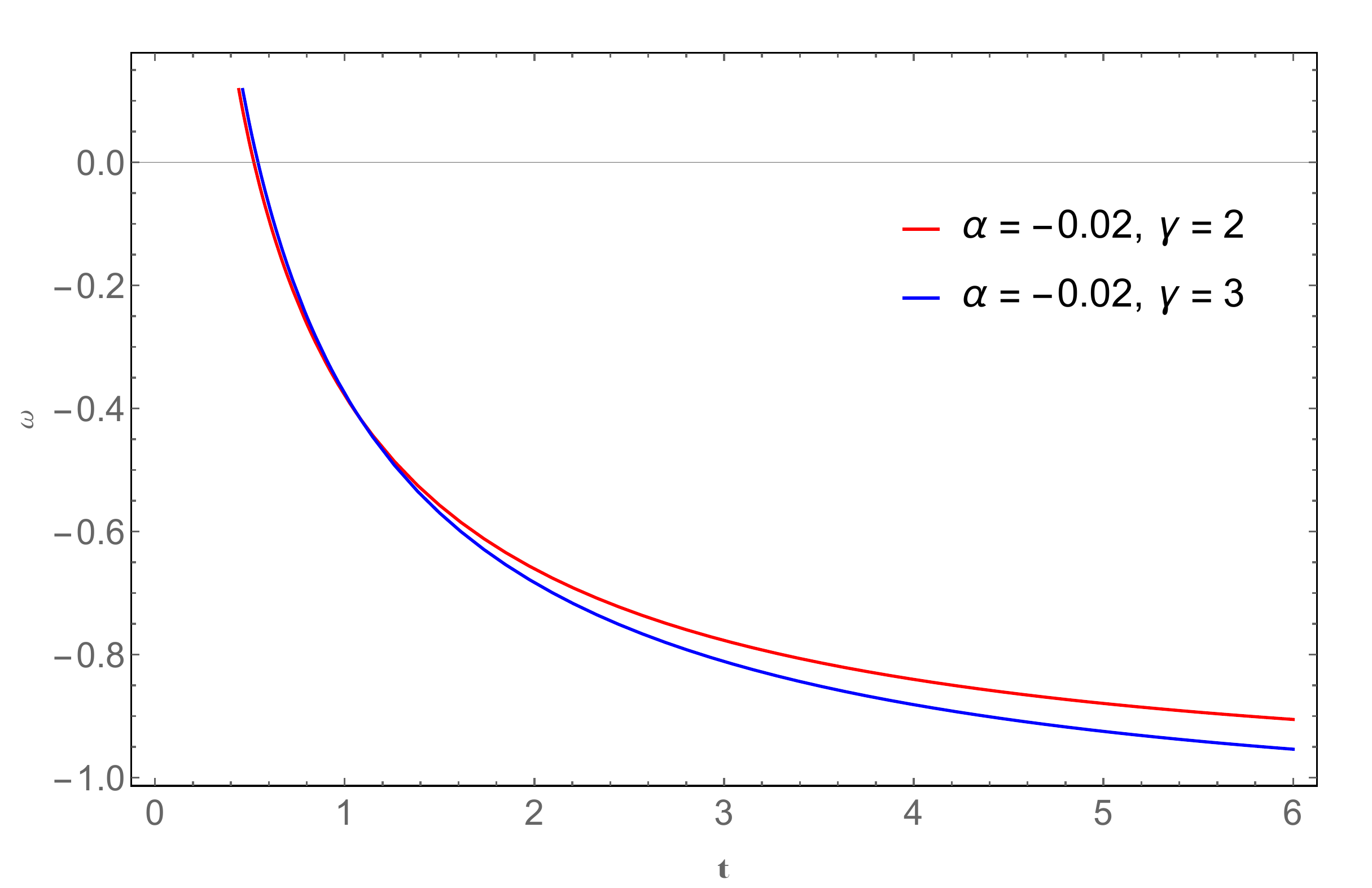}
\caption{Evolution of $\protect\omega$ with time.}\label{fig5}
\end{figure}

The evolution of the energy density, pressure and corresponding equation of state (EoS) parameter $\omega=p/\rho$ with $%
m=0.27, n=0.75$ are shown in Figures \ref{fig3}-\ref{fig5}, in which the time units are Gyr. 

\section{Energy conditions}\label{sec:ec}

Energy condition (ECs), in the context of a wide class of covariant theories including GR,
are relations one demands for the energy-momentum tensor of matter to satisfy in
order to try to capture the idea that ``energy should be positive". By imposing so, one can obtain constraints to the free parameters of the concerned model.

The standard point-wise ECs are (\cite{Hawking/1973, Wald/1984, Visser/1995}): \\
Null energy condition (NEC): $\rho+p \geq 0$;\\
Weak energy condition (WEC): $\rho \geq 0$, \ \ \ $\rho+p \geq 0$;\\
Strong energy condition (SEC): $\rho+3p \geq 0$;\\  
Dominant energy condition (DEC): $\rho \geq \mid p \mid$.\\
Generally the above ECs are formulated from the Raychaudhuri equation, which
describes the behavior of space-like, time-like or light-like curves in gravity. 

In the present model the energy-momentum tensor has the form of a perfect fluid. So, we will use
the above relations for analysing the ECs in $f(R,T)$ theory. The behaviour of the ECs with $m=0.27, n=0.75$ and $\alpha=-0.02$ are given in Figures \ref{fig10}-\ref{fig13} below.

\begin{figure}[h]
\centering
\includegraphics[width=70mm]{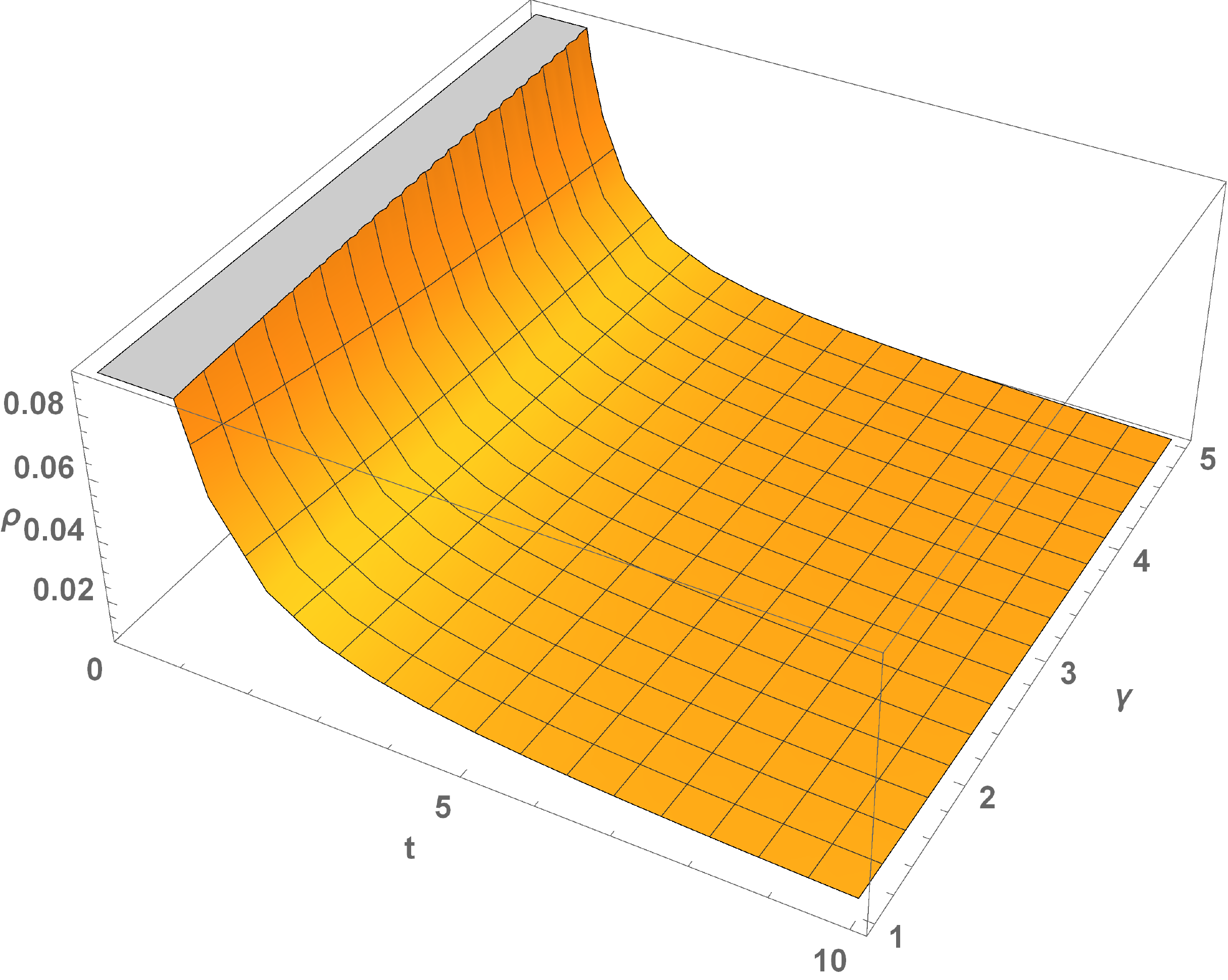}
\caption{Evolution of $\protect\rho$.}\label{fig10}
\end{figure}

\begin{figure}[h!]
\centering
\includegraphics[width=70mm]{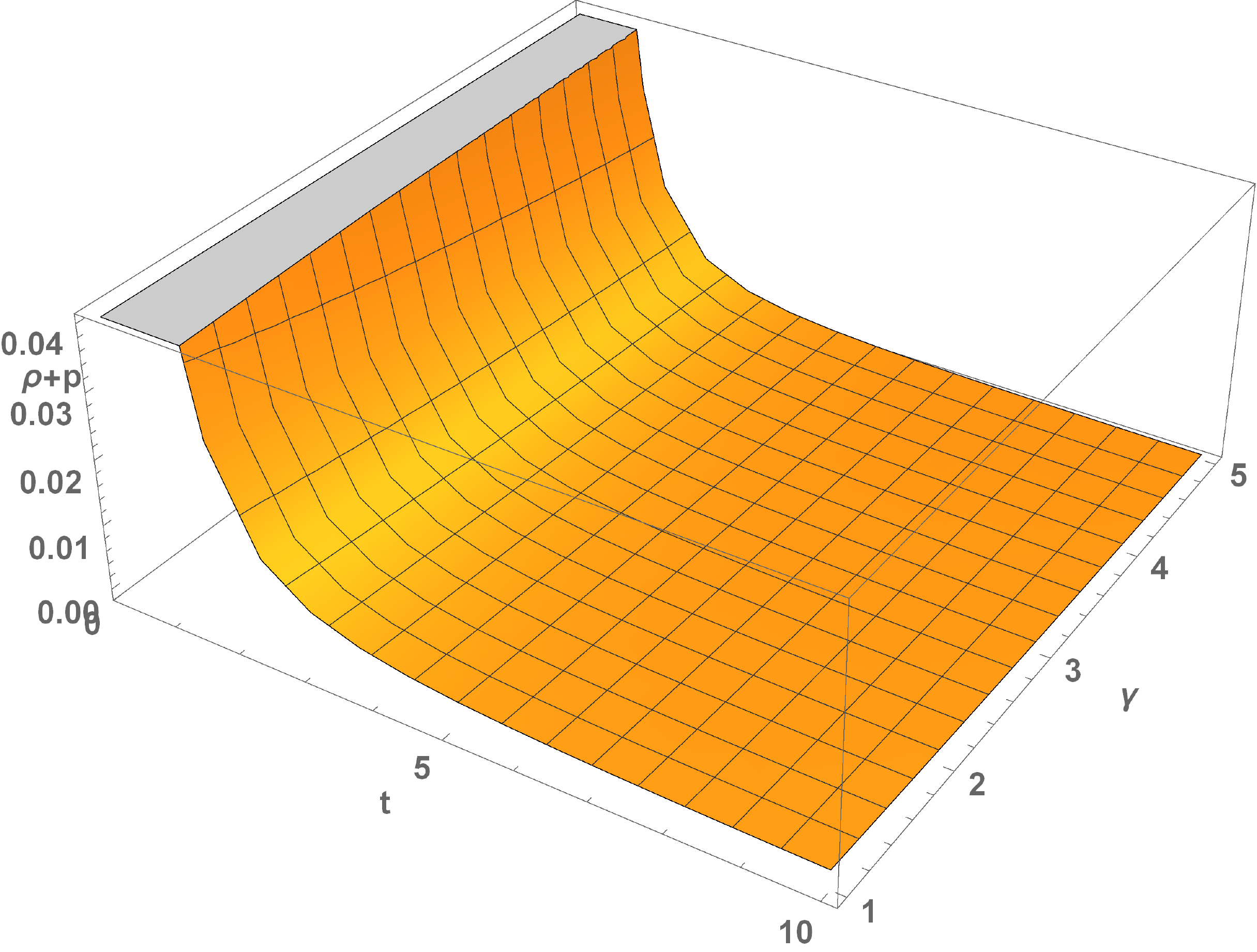}
\caption{Validation of NEC.}\label{fig11}
\end{figure}

\begin{figure}[h!]
\centering
\includegraphics[width=70mm]{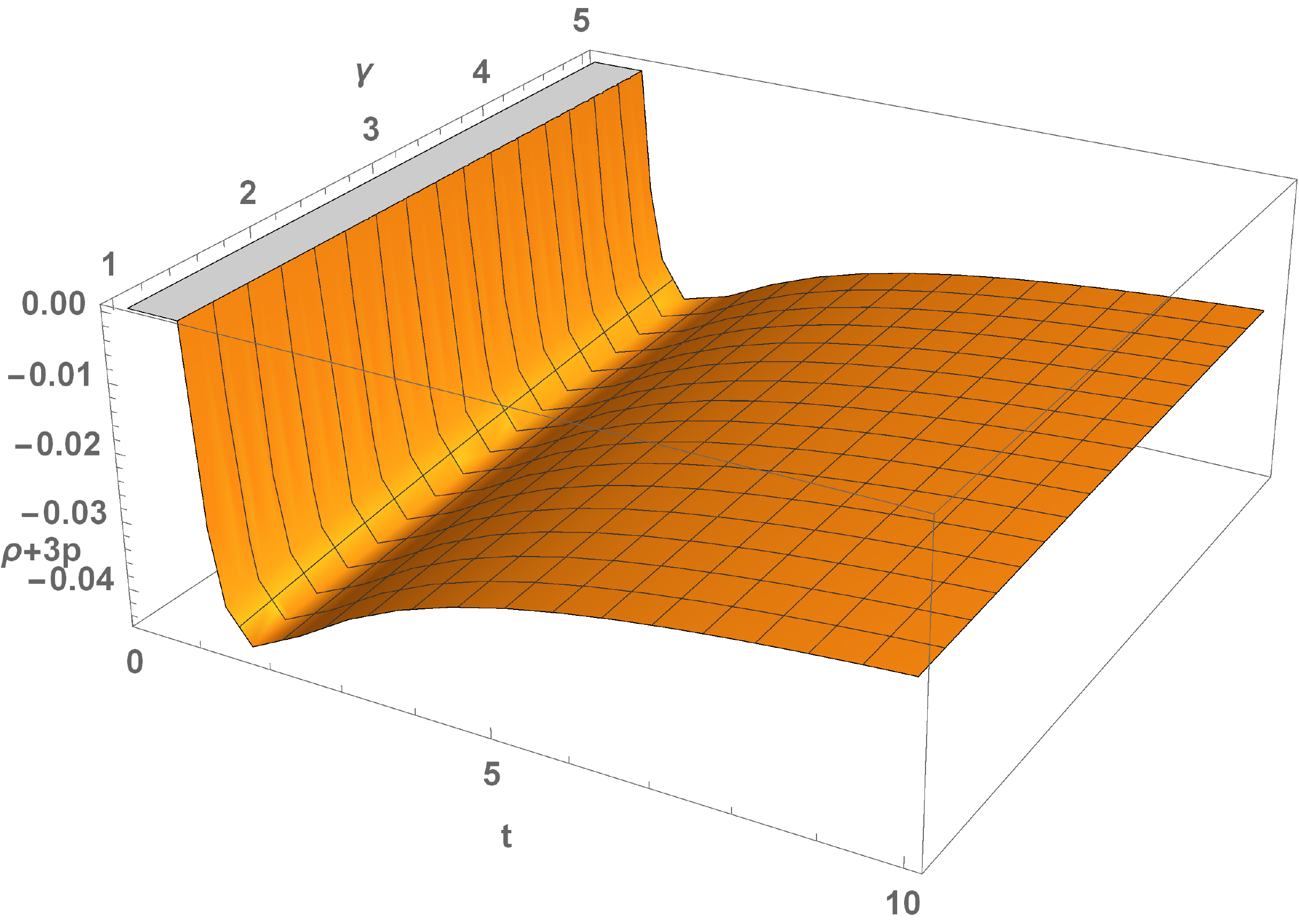}
\caption{Violation of SEC.}\label{fig12}
\end{figure}

\begin{figure}[h!]
\centering
\includegraphics[width=70mm]{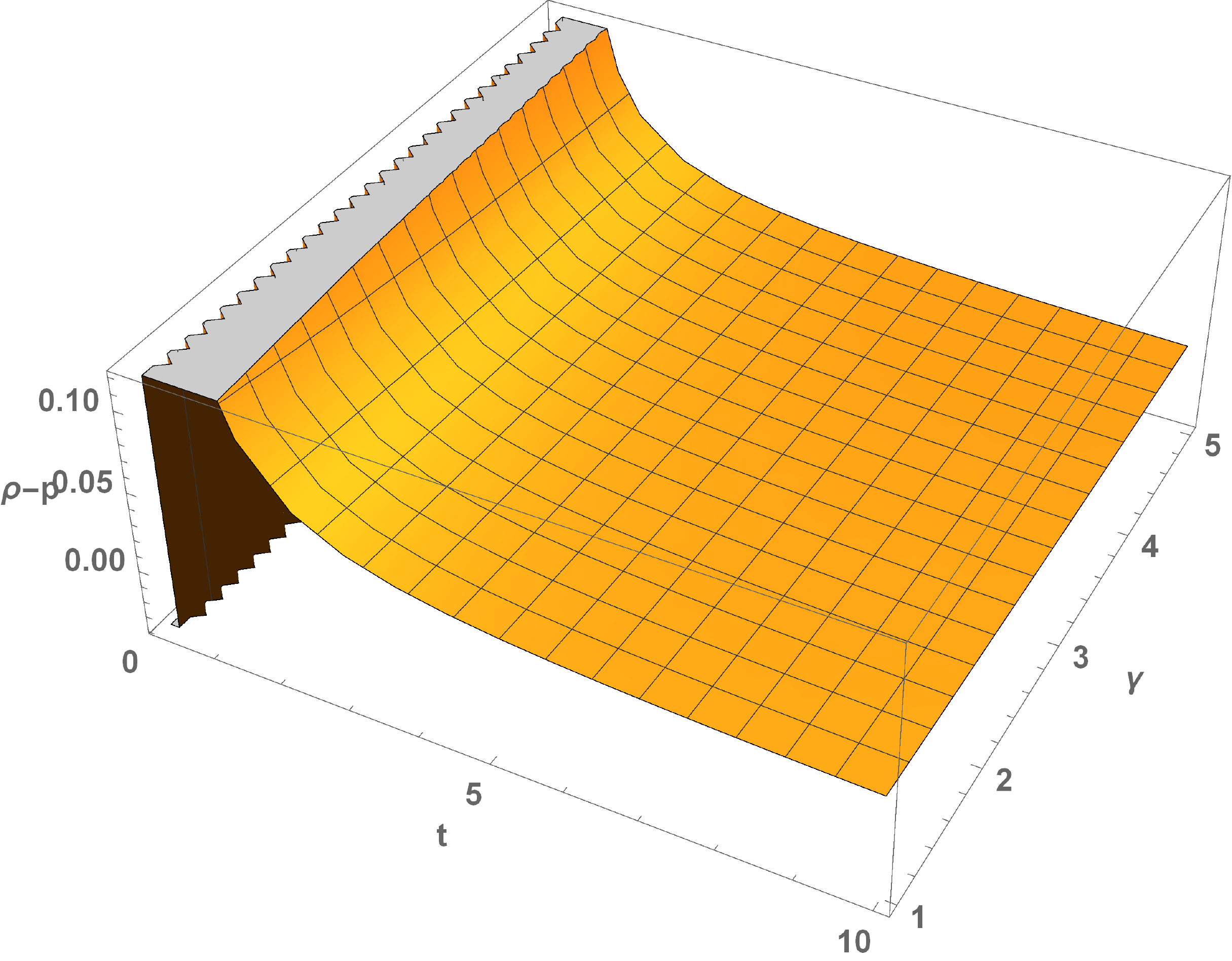}
\caption{Validation of DEC.}\label{fig13}
\end{figure}

\section{Discussion and conclusions}

In this article we have proposed an $f(R,T)$ cosmological model. For the functional form of the function $f(R,T)$, we investigated a quadratic correction to $R$, as in the well-known SM, together with a linear term on $T$. The substitution of such an $f(R,T)$ function in the gravitational action generates extra terms in the field equations (\ref{frt2}) and, consequently, in the Friedmann-like equations (\ref{00}), (\ref{11}). Those extra terms bring some new informations regarding the dynamics of the universe, as we discuss below.

Our solution for the scale factor $a(t)$ is a hybrid expansion law, well described in \cite{akarsu/2014}. From (\ref{14}), we have obtained the cosmological parameters, namely Hubble factor and deceleration parameter. Specifically, for the deceleration parameter, we could separate it in two phases: one describing the decelerated and the other describing the accelerated regime of the universe expansion. This can be well-checked in Figure \ref{fig2}, in which the transition redshift between these two stages agrees with observational data. Moreover, one can also note that, remarkably, the present ($z=0$) values for $q$, named $-0.178$ for $m=0.27$ and $-0.157$ for $m=0.25$, are also in agreement with observational data (\cite{hinshaw/2013}).

We have also obtained solutions for the material content of the universe, named $\rho$ and $p$ (Eqs.(\ref{rhot})-(\ref{um})). In Figs.\ref{fig3}-\ref{fig5} we plot the evolution of $\rho$, $p$ and $\omega$, the EoS parameter. The values chosen for $\alpha$ and $\gamma$ respect the energy conditions outcomes presented in Section \ref{sec:ec}. In Fig.\ref{fig4} we see that the pressure of the universe starts in positive values and then assumes negative values. In standard model of cosmology, a negative pressure fluid is exactly the mechanism responsible for accelerating the universe expansion. In the present model, such a behavior for the pressure was naturally obtained. 

It is also worth stressing that the EoS parameter shows a transition from a decelerated to an accelerated regime of the expansion of the universe. This can well be seen in Fig.\ref{fig5}, by recalling that from standard cosmology, the latter regime may happen if $\omega<-1/3$  (\cite{ryden/2003}). Moreover, it can be seen from such a figure that as time passes by, $\omega\rightarrow-1$, in accordance with recent observational data on the cosmic microwave background temperature fluctuations (\cite{hinshaw/2013}).

Furthermore, in Figs.\ref{fig10}-\ref{fig13} we plotted the ECs from the material solutions of our cosmological model. Those  figures were plotted in terms of $\gamma$, for fixed $\alpha=-0.02$. They show the validation of WEC, NEC and DEC, with a wide range of acceptable values for $\gamma$.

On the other hand, Fig.\ref{fig12} shows violation of SEC. However, as it has been deeply discussed by \cite{barcelo/2002} an early and late-time accelerating universe must violate SEC.

As a further work, we can look for the $f(R,T)=R+\alpha R^{2}+2\gamma T$ gravity universe at very early times, particularly investigating the production of scalarons in this model. \cite{ms/2016} have shown that the trace of the energy-momentum tensor contribution in the theory is higher for the early universe when compared to the late-time contribution. According to \cite{starobinsky/2007}, some mechanism should work in the early universe to prohibit the scalaron overproduction within SM. The high contribution of the terms proportional to $T$ in the early universe may well be this mechanism.

\bigskip \textbf{Data Availability}
The data used to support the fndings of this study are
included within the article.

\bigskip \textbf{Conflicts of Interest}
The author declares that they have no conficts of interest.

{\bf Acknowledgements}

PHRSM thanks S\~ao Paulo Research Foundation (FAPESP), grant 2015/08476-0, for financial support. PKS acknowledges DST, New Delhi, India for providing facilities through DST-FIST lab, Department of Mathematics, where a part of this work was done. The authors are also thankful to the anonymous referees, whose valuable comments have helped to improve the standard of manuscript. RACC is partially supported by FAPESP (Foundation for Support to Research of the State of S\~ao Paulo) under grants numbers 2016/03276-5 and 2017/26646-5.


\begin{thebibliography}{}

\bibitem[Albert et al. (2017)]{albert/2017} Albert, A. et al., 2017, Phys. Dark Univ., 16, 49
\bibitem[Alves et al. (2016)]{amam/2016} Alves, M. E. S., Moraes, P. H. R. S.,  Ara\'ujo, J. C. N. \& Malheiro, M., 2016, Phys. Rev. D, 94, 024032
\bibitem[Antoniadis et al. (2013)]{antoniadis/2013} Antoniadis, J. et al.,   2013, Science, 340, 448
\bibitem[Astashenok et al. (2015)]{astashenok/2015} Astashenok, A. V. et al., 2015, Phys. Lett. B, 742, 160
\bibitem[Baffou et al. (2017)]{baffou/2017} Baffou, E. H., Salako, I. G. \& Houndjo, M. J. S., 2017, Int. J. Geom. Meth. Mod. Phys., 14, 1750051
\bibitem[Barcelo \& Visser (2002)]{barcelo/2002} Barcelo, C. \& Visser, M., 2002, Int. J. Mod. Phys. D, 11, 1553
\bibitem[Barrientos et al. (2018)]{barrientos/2018} Barrientos, E. et al. 2018, Phys. Rev. D, 97, 104041
\bibitem[Basilakos et al. (2013)]{basilakos/2013} Basilakos, S. et al., 2013, Phys. Rev. D, 87, 123529
\bibitem[Behrouz et al. (2017)]{behrouz/2017} Behrouz, N. et al.,  2017, Phys. Dark Univ., 15, 72
\bibitem[Bull et al. (2016)]{bull/2016} Bull, P. et al.  2016, Phys. Dark Univ., 12, 56
\bibitem[Capozziello et al. (2005)]{capozziello/2005} Capozziello, S. et al., 2005, Phys. Rev. D, 71, 043503
\bibitem[Capozziello et al. (2007)]{capozziello/2007b} Capozziello, S. et al., 2007, Phys. Rev. D, 76, 104019
\bibitem[Capozziello et al. (2014)]{Capozziello14} Capozziello, S., Luongo, O. \& Saridakis, E. N., 2014, Phys. Rev. D, 90, 044016
\bibitem[Capozziello et al. (2015)]{Capozziello15} Capozziello, S., Farooq, O., Luongo, O. \& Ratra, B., 2015, Phys. Rev. D, 91, 124037
\bibitem[Chiba (2003)]{chiba/2003} Chiba, T., 2003, Phys. Lett. B, 575, 1
\bibitem[Chiba et al. (2007)]{chiba/2007} Chiba, T. et al., 2007, Phys. Rev. D, 75, 124014
\bibitem[Correa et al. (2015)]{cmsdr/2015} Correa, R. A. C., Moraes, P. H. R. S., de Souza Dutra, A. a\& da Rocha, R., 2015, Phys. Rev. D, 92, 126005
\bibitem[Correa \& Moraes (2016)]{cm/2016} Correa, R. A. C. \& Moraes, P. H. R. S., 2016, Eur. Phys. J. C, 76, 100
\bibitem[Das et al. (2016)]{das/2016} Das, A., Rahaman, F. and Guha, B. K. \& Ray, S., 2016, Eur. Phys. J. C, 76, 654
\bibitem[Felice \& Tsujikawa (2010)]{de_felice/2010} de Felice, A. \& Tsujikawa, S.  2010, Liv. Rev. Rel., 13, 161
\bibitem[Demorest et al. (2010)]{demorest/2010} Demorest, P. B. et al.,  2010, Nature, 467, 1081
\bibitem[Dil (2017)]{dil/2017} Dil, E.,  2017, Phys. Dark Univ., 16, 1
\bibitem[Dolgov \& Kawasaki (2003)]{dolgov/2003} Dolgov, A. D. \& Kawasaki, M., 2003, Phys. Lett. B, 573, 1
\bibitem[Erickcek et al. (2006)]{erickcek/2006} Erickcek, A. L. et al., 2006, Phys. Rev. D, 74, 121501
\bibitem[Evslin (2016)]{evslin/2016} Evslin, J., 2016, Phys. Dark Univ., 13, 126
\bibitem[Farooq et al. (2017)]{Farooq17} Farooq, O., Madiyar, F., Crandall, S. \& Ratra, B., 2017, Astrophys. J., 835, 26
\bibitem[Fradkin \& Tseytlin (1985)]{fradkin/1985} Fradkin, E. S. \& Tseytlin, A. A., 1985, Nuclear Phys. B, 261, 1
\bibitem[Friedan et al. (1986)]{friedan/1986} Friedan, D. et al., 1986, Nuclear Phys. B, 271, 93
\bibitem[Germani (2017)]{germani/2017} Germani, C., 2017, Phys. Dark Univ., 15, 1
\bibitem[Harko et al. (2011)]{harko/2011} Harko, T. et al., 2011, Phys. Rev. D, 84, 024020
\bibitem[Hinshaw et al. (2013)]{hinshaw/2013} Hinshaw, G. et al., 2013, Astrophys. J. Supp., 208, 19
\bibitem[Hawking \& Ellis (1973)]{Hawking/1973} Hawking, S. W. \& Ellis, G. F. R.,  1973, The large scale structure of spacetime, Cambridge University Press, England
\bibitem[Jennen \& Pereira (2016)]{jennen/2016} Jennen, H. \& Pereira, J. G., 2016, Phys. Dark Univ., 11, 49
\bibitem[Liu et al. (2016)]{liu/2016} Liu, Z.-E et al.,  2016, Phys. Dark Univ., 14, 21
\bibitem[Moraes (2015)]{moraes/2015} Moraes, P. H. R. S., 2015, Eur. Phys. J. C, 75, 168
\bibitem[Moraes et al. (2016)]{mrc/2016} Moraes, P. H. R. S., Ribeiro, G. \& Correa, R. A. C., 2016, Astrophys. Space Sci., 361, 227
\bibitem[Moraes \& Santos (2016)]{ms/2016} Moraes, P. H. R. S. \& Santos, J. R. L., 2016, Eur. Phys. J. C, 76, 60
\bibitem[Moraes et al. (2016)]{mam/2016} Moraes, P. H. R. S., Arba\~nil, J. D. V. \& Malheiro, M., 2016, JCAP, 06, 005
\bibitem[Moraes \& Sahoo (2017)]{ms/2017} Moraes, P. H. R. S. \& Sahoo, P. K., 2017, Eur. Phys. J. C, 77, 480
\bibitem[Moraes \& Sahoo (2017)]{ms/2017b} Moraes, P. H. R. S. \& Sahoo, P. K., 2017, Phys. Rev. D, 96, 044038
\bibitem[Moraes et al. (2017)]{mcl/2017} Moraes, P. H. R. S., Correa, R. A. C. \& Lobato, R. V., 2017, JCAP, 07, 029
\bibitem[Myrzakulov (2012)]{myrzakulov/2012} Myrzakulov, R., 2012, Eur. Phys. J. C, 72, 2203
\bibitem[Nojiri \& Odintsov (2006)]{nojiri/2006} Nojiri, S. \& Odintsov, S. D., 2006, Phys. Rev. D, 74, 086005
\bibitem[Nojiri \& Odintsov (2011)]{nojiri/2011} Nojiri, S. \& Odintsov, S. D., 2011, Phys. Rept., 505, 59
\bibitem[Nojiri et al. (2017)]{nojiri/2017} Nojiri, S., Odintsov, S. D. \& Oikonomou, V.K.,  2017, Phys. Rept., 692, 1
\bibitem[Noureen \& Zubair (2015)]{noureen/2015} Noureen, I. \& Zubair, M., 2015, Eur. Phys. J. C, 75, 62
\bibitem[Noureen et al. (2015)]{noureen/2015b} Noureen, I. et al., 2015, Eur. Phys. J. C, 75, 323
\bibitem[Olmo (2005)]{olmo/2005} Olmo, G. J., 2005, Phys. Rev. D, 72, 083505
\bibitem[Olmo (2007)]{olmo/2007} Olmo, G. J., 2007, Phys. Rev. D, 75, 023511
\bibitem[Ozgur et al. (2014)]{akarsu/2014} Ozgur, A. et al., 2014, JCAP, 01, 022
\bibitem[Padmanabhan (2003)]{padmanabhan/2003} Padmanabhan, T.  2003, Phys. Rep., 380, 235
\bibitem[Perlmutter et al. (1999)]{perlmutter/1999} Perlmutter, G. et al., 1999, Astrophys. J., 517, 565
\bibitem[Reddy et al. (2012)]{reddysn/2012} Reddy, D. R. K., Santikumar, R.\& 
Naidu, R. L., 2012, Astrophys. Space Sci., 342, 249
\bibitem[Resco et al. (2016)]{resco/2016} Resco, M. A. et al., 2016, Phys. Dark Univ., 13, 147
\bibitem[Riess et al. (1988)]{riess/1998} Riess, A. G. et al., 1988, Astron. J., 116, 1009
\bibitem[Riess et al. (1998)]{riess/1998} Riess, A. G. et al., 1998, Astron. J., 116, 1009
\bibitem[Rinaldi (2017)]{rinaldi/2017} Rinaldi., M., 2017, Phys. Dark Univ., 16, 14
\bibitem[Ryden (2003)]{ryden/2003} Ryden, B., 2003, Introduction to Cosmology, Addison Wesley, San Francisco
\bibitem[Shabani \& Farhoudi (2014)]{shabani/2014} Shabani, H. \& Farhoudi, M., 2014, Phys. Rev. D, 90, 044031
\bibitem[Sharif \& Yousaf (2014)]{sy/2014} Sharif, M. \& Yousaf, Z., 2014, Astrophys. Space Sci., 354, 471
\bibitem[Sharif \& Siddiqa (2017)]{sharif/2017} Sharif, M. \& Siddiqa, A., 2017, Phys. Dark Univ., 15, 105
\bibitem[Singh \& Singh (2015)]{singh/2015} Singh, V. \& Singh, C. P., 2015, Int. J. Theor. Phys., 55, 1257
\bibitem[Sotiriou \& Faraoni (2010)]{sotiriou/2010} Sotiriou, T. P. \& Faraoni, V.  2010, Rev. Mod. Phys., 82, 451
\bibitem[Starobinsky (1980)]{starobinsky/1980} Starobinsky, A. A., 1980, Phys. Lett., 91B, 99
\bibitem[Starobinsky (2007)]{starobinsky/2007} Starobinsky, A. A., 2007, JETP Lett., 86, 157
\bibitem[Tsujikawa (2008)]{tsujikawa/2008} Tsujikawa, S., 2008, Phys. Rev. D, 77, 023507
\bibitem[Visser (1995)]{Visser/1995} Visser, M., 1995, Lorentzian wormholes, AIP Press, New York   
\bibitem[Wald (1984)]{Wald/1984} Wald, R. M., 1984, General Relativity, University of Chicago Press, Chicago
\bibitem[Witten (1986)]{witten/1986} Witten, E., 1986, Nuclear Phys. B, 268, 253
\bibitem[Wu et al. (2018)]{wu/2018} Wu, J., Li, G., Harko, T. et al., 2018, Eur. Phys. J. C, 78, 430
\bibitem[Yousaf et al. (2017)]{yousaf/2017b} Yousaf, Z. et al., 2017, Mod. Phys. Lett. A, 32, 1750163
\bibitem[Yousaf et al. (2017)]{yousaf/2017} Yousaf, Z. et al., 2017, Eur. Phys. J. Plus, 132, 268
\bibitem[Zaregonbadi et al. (2016)]{zaregonbadi/2016} Zaregonbadi, R. et al., 2016, Phys. Rev. D, 94, 084052
\bibitem[Zhang (2017)]{zhang/2017} Zhang, Y., 2017, Phys. Dark Univ., 15, 82
\bibitem[Zubair \& Noureen (2015)]{zubair/2015} Zubair, M. \& Noureen, I., 2015, Eur. Phys. J. C, 75, 265
\bibitem[Zubair et al. (2016)]{zubair/2016} Zubair, M. et al., 2016, Eur. Phys. J. C, 76, 444
\bibitem[Zubair et al. (2018)]{zubair/2018} Zubair, M. et al., 2018, Symmetry, 10, 463

\end{thebibliography}
\end{document}